\documentclass[fleqn,usenatbib]{mnras}

\usepackage{newtxtext,newtxmath}

\usepackage[T1]{fontenc}

\DeclareRobustCommand{\VAN}[3]{#2}
\let\VANthebibliography\thebibliography
\def\thebibliography{\DeclareRobustCommand{\VAN}[3]{##3}\VANthebibliography}

\usepackage{graphicx}	
\usepackage{amsmath}	

\makeatletter
\DeclareRobustCommand{\HI}{%
  \mbox{H\check@mathfonts\fontsize\sf@size\z@\selectfont I~}%
}
\makeatother

\makeatletter
\DeclareRobustCommand{\H2}{%
  \mbox{H\check@mathfonts\fontsize\sf@size\z@\selectfont 2~}%
}
\makeatother

\title[Black hole feedback and stellar bars]{Supermassive black hole feedback quenches disc galaxies and suppresses bar formation in {\sc{TNG50}}}

\author[M. Frosst]{
Matthew Frosst$^{1,2}$\thanks{E-mail: matt.frosst@icrar.org},
Danail Obreschkow$^{1,2}$,
Aaron Ludlow$^{1,2}$,
Connor Bottrell$^{1}$, and
Shy Genel$^{3,4}$
\\
$^{1}$International Centre for Radio Astronomy Research (ICRAR), University of Western Australia, Crawley, WA 6009, Australia\\
$^{2}$ARC Centre of Excellence for All Sky Astrophysics in 3 Dimensions (ASTRO 3D)\\
$^{3}$Center for Computational Astrophysics, Flatiron Institute, 162 5th Avenue, New York, NY 10010, USA\\
$^{4}$Columbia Astrophysics Laboratory, Columbia University, 550 West 120th Street, New York, NY 10027, USA
}

\date{Accepted XXX. Received YYY; in original form ZZZ}

\pubyear{2023}

\begin{document}
\label{firstpage}
\pagerange{\pageref{firstpage}--\pageref{lastpage}}
\maketitle

\begin{abstract}
We use the cosmological magneto-hydrodynamical simulation \textsc{TNG50} to study the relationship between black hole feedback, the presence of stellar bars, and star formation quenching in Milky Way-like disc galaxies. Of our sample of 198 discs, about $63$ per cent develop stellar bars that last until $z=0$. After the formation of their bars, the majority of these galaxies develop persistent 3-15 kpc wide holes in the centres of their gas discs. Tracking their evolution from $z=4$ to $0$, we demonstrate that barred galaxies tend to form within dark matter haloes that become centrally disc dominated early on (and are thus unstable to bar formation) whereas unbarred galaxies do not; barred galaxies also host central black holes that grow more rapidly than those of unbarred galaxies. As a result, most barred galaxies eventually experience kinetic wind feedback that operates when the mass of the central supermassive black hole exceeds ${\rm M_{BH}} \gtrsim 10^8\, {\rm M_{\odot}}$. This feedback ejects gas from the central disc into the circumgalactic medium and rapidly quenches barred galaxies of their central star formation. If kinetic black hole feedback occurs in an unbarred disc it suppresses subsequent star formation and inhibits its growth, stabilising the disc against future bar formation. Consequently, most barred galaxies develop black hole-driven gas holes, though a gas hole alone does not guarantee the presence of a stellar bar. This subtle relationship between black hole feedback, cold gas disc morphology, and stellar bars may provide constraints on subgrid physics models for supermassive black hole feedback. 
\end{abstract}



\section{Introduction}
In the $\Lambda$CDM cosmology, the dark matter density field of the universe is nearly scale-free, but the galaxies embedded within it are not. 
One of the most significant transitions in galaxy properties occurs near the mass of the Milky Way (MW), corresponding to a halo mass of $\sim 10^{12}\, {\rm M_{\odot}}$. At this scale, the galaxy population gradually transitions from star-forming discs to quiescent spheroids \citep[e.g.][]{Moffett2016}. 
Intriguingly, even within the disc population substantial changes occur around this mass scale. 
For example, more massive galaxies become increasingly bulge-dominated \citep{Bell2017}, the incidence of stellar bars may increase \citep{Melvin2014, Gavazzi2015, Erwin2018}, and cold gas deficits that reduce star formation (SF) activity become more common \citep[e.g.][]{Kauffmann2003}. 
Moreover, massive discs interact more strongly with their central supermassive black holes (SMBHs), which become increasingly active, as demonstrated by the rising fraction of Seyfert galaxies above this mass scale \citep{Sampaio2023}. 

While the observational evidence for these transitions is robust, the underlying mechanisms that cause them remain under debate. 
Specifically, the relationships between bars, SMBH feedback, and the quenching of SF are not yet fully understood. 
Accreting SMBHs require gas to be transported to the disc centre as fuel, with torques from bars being a possible driver \citep{Maciejewski2004, Jogee2006, Fanali2015}. 
However, bar driven SMBH fueling remains controversial: some observations suggest that SMBH activity is enhanced by the presence of a bar \citep{Knapen2000, Laine2002, SilvaLima2022, Garland2024}, while others do not \citep{Lee2012, Goulding2017, Zee2023}. 

From a theoretical standpoint, simulating the interplay between stellar bars, SMBHs, and SF is difficult to study because it demands high-resolution simulations that incorporate uncertain sub-grid models to capture complex physics below the resolution limit. 
Simulations provided the first compelling evidence that SMBH feedback plays a crucial role in quenching the population of massive galaxies \citep[e.g.][]{Springel2005, Bower2006, Dubois2013, Weinberger2018}. 
More recently, simulations have been used to quantify the importance of SMBH feedback on the properties of gas in galaxies \citep[e.g.,][]{Davies2020, Zinger2020, Ramesh2023, Wellons2023}, and to constrain SMBH models \citep[e.g.][]{Genel2014}. 
However, despite significant effort, the mechanisms by which energy from SMBH feedback couples to the gas, as well as the importance of other heating mechanisms, are not yet understood \citep[e.g.][]{McNamara2007,Yuan2014}. 

Stellar bars are ubiquitous among galaxies in the local universe \citep{Masters2011}, and are thus frequently studied using cosmological simulations \citep[e.g.][]{BlazquezCalero2020, Ansar2023, Fragkoudi2024}. Simulated barred galaxies tend to be gas poor and quenched, as expected \citep[][]{Algorry2017, RosasGuevara2020, RosasGuevara2022, Zhu2020}. 
Furthermore, \citet{Zana2019} showed that SMBH feedback can influence the properties of bars, for example causing them to be shorter and weaker than they otherwise would be \citep[see also][]{Irodotou2022, Zhou2020,Semczuk2024}, though the interactions between bars and SMBHs remain unclear. 

However, robustly simulating the formation of stellar bars in disc galaxies is not trivial. 
Although it is not difficult to produce bars in simulations of disc-dominated systems \citep{Hohl1971, OP1973, Athanassoula1986}, the challenge lies in generating populations of barred galaxies with converged quantitative properties, such as their incidence, formation time, length, and pattern speed. 
For instance, convergence studies using idealised simulations show that including live dark matter haloes that can interact with the bar is crucial, and that simulations require $\sim10^6$ disc particles and $\sim10^7$ dark matter halo particles to achieve reliable results \citep[e.g.][]{Dubinski2009, Frosst2024}. These requirements challenge cosmological hydrodynamical simulations, which typically produce galaxies with far fewer stellar and dark matter particles. 

The aim of this study is to explore the connection between SMBH feedback, stellar bars, and SF quenching in disc galaxies using a state-of-the-art cosmological simulation from the IllustrisTNG project (hereafter TNG). Specifically, we focus on a carefully selected and well-studied sample of 198 MW- and M31-like disc galaxies from TNG50 \citep[see][]{Pillepich2023}. This sample allows us to simultaneously focus on the mass scale where the aforementioned scale-dependent features of disc galaxies become significant and to reach the high resolutions required to reliably model bars properties. Additionally, this disc sample consists of galaxies in isolated environments (at $z = 0$), enabling us to focus primarily on the internal processes driving secular bar formation. 

This paper is organised as follows. In Section~\ref{sec:methods}, we describe the \textsc{TNG50} simulation and the disc sample that we use for our analysis; we then introduce various definitions and analysis techniques. In Section~\ref{sec:results} we present the results of our analysis and explain the links between SMBH feedback, central gas disc morphology, and the formation of stellar bars.  Finally, we summarise our results in Section~\ref{sec:conclusions}. 

\begin{figure*}
	\includegraphics[width=\textwidth]{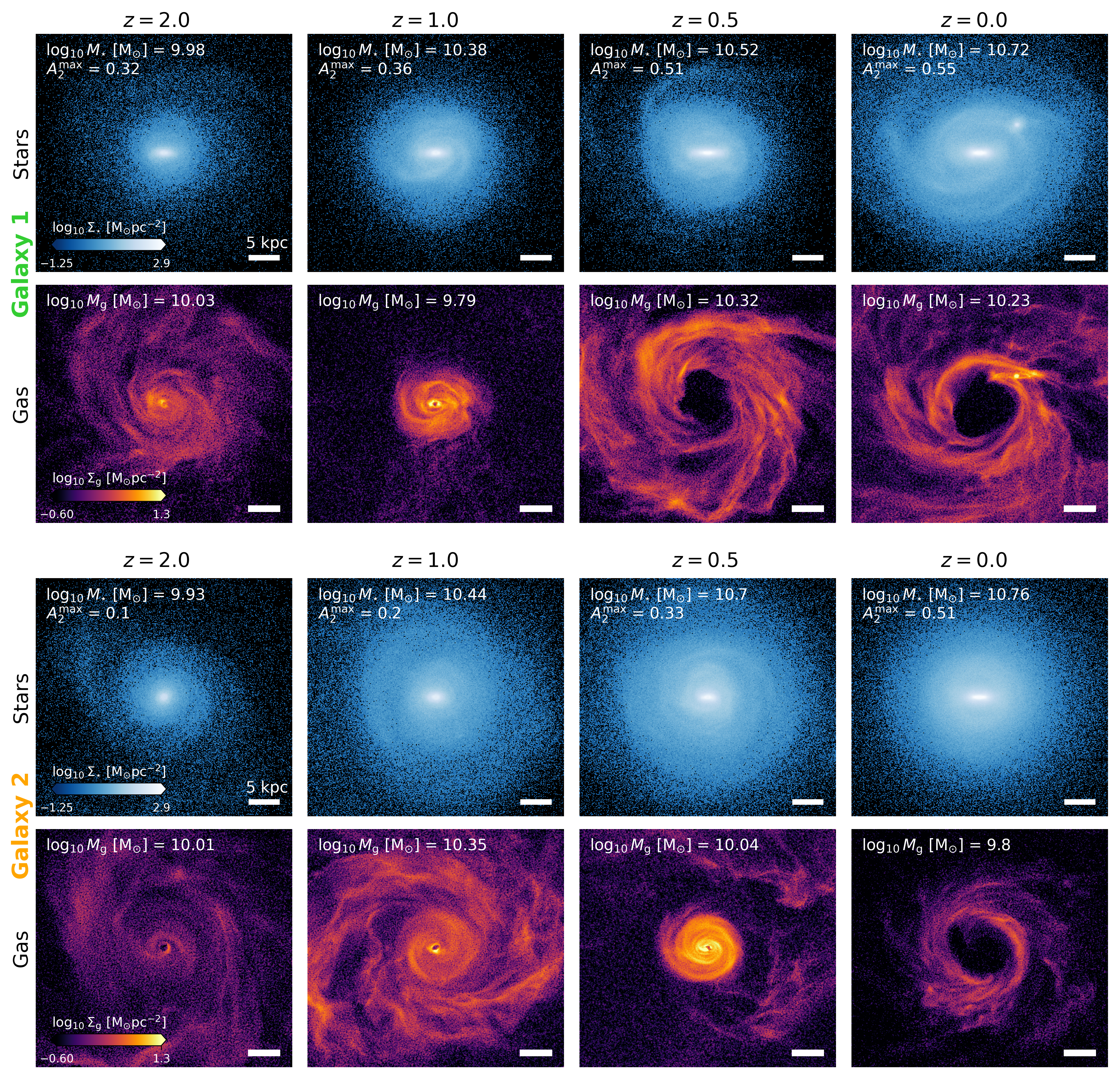}
    \caption{The different panels show shapshots of the time evolution from $z=2$ to $0$ (left to right) of two example galaxies: subfindID $530852$ (Galaxy 1) is shown in the upper two rows, and subfindID $534628$ (Galaxy 2) in the lower two rows. 
    For each galaxy, we show the face-on stellar surface mass density ($\Sigma_{\star}$; top rows, blue colours) and gas surface mass density ($\Sigma_{\rm g}$; bottom rows, red colours). 
    In the top left corner of the $\Sigma_{\star}$ panels we label the stellar mass within $r \leq 30$ kpc and bar strength. 
    Similarly, in the top left corner of the $\Sigma_{\rm g}$ panels we label the gas mass within $r \leq 30$ kpc. 
    The white lines in the bottom right of each panel have a length of $5$ kpc.}
    \label{fig:projection}
\end{figure*}

\section{Methods}\label{sec:methods}
For our analysis, we adopt a cylindrical coordinate system whose $z$-axis is aligned with the total angular momentum vector of stellar particles and star-forming gas cells within two stellar half mass radii, $r_{\star,1/2}$. 
In this coordinate system, $r = (R^2 + z^2)^{1/2}$ is the 3D radial coordinate, where $R=(x^2+y^2)^{1/2}$ is the distance from the $z$-axis, and $z$ is the height above the disc. 

\subsection{The \textsc{TNG50} simulation}\label{sec:code}
We focus on the \textsc{TNG50} simulation \citep[TNG50-1;][]{Pillepich2019, Nelson2019a, Nelson2019b}, which is the highest resolution simulation of the TNG project \citep[][]{Marinacci2018, Naiman2018, Nelson2018, Springel2018, Pillepich2018a, Nelson2019a}. 
The simulation was run with the \textsc{AREPO} moving-mesh code \citep{Springel2010}, using cosmological parameters from \citet{Planck2016}. 

\textsc{TNG50} follows the evolution of gas cells, dark matter (DM), black hole and stellar particles, and magnetic fields from $z=127$ to $z=0$ in a $51.7$ cMpc volume (note that length units prefixed with a "c" are co-moving). 
The simulation initially has $2\times2160^3$ mass elements, half of which are DM particles with mass $4.5\times10^5\, {\rm M_{\odot}}$ and the other half are gas cells with a target mass of $8.5\times10^4\, {\rm M_{\odot}}$. 
The Plummer equivalent gravitational softening length of gas elements is dynamically adapted to the effective cell radius, but has a minimum value of $\epsilon_{\rm g} = 72$ pc. 
Collisionless particles (DM and stars) have a Plummer softening length of $\epsilon_{\rm c} = 575$ cpc until $z=1$, below which they have a fixed physical softening length of $\epsilon_{\rm c} = 288$ pc. 

\textsc{TNG50} includes models for star formation, stellar feedback, active galactic nuclei (AGN) feedback, supermassive black hole growth, stellar evolution, and chemical enrichment \citep[see][for details]{Pillepich2018b, Springel2018, Nelson2018, Naiman2018, Marinacci2018}. Details about the subgrid physics and calibration process can be found in \citet{Weinberger2017} and \citet{Pillepich2018a}. 

Galaxies and DM haloes were identified using the \textsc{SUBFIND} algorithm \citep{Springel2001} and linked between snapshots by the \textsc{Sublink} merger tree code \citep{RodriguezGomez2015}. \textsc{SUBFIND} also returns a number of basic properties of haloes, for instance, $M_{\rm 200c}$, the mass within the radius $r_{\rm 200c}$ that encloses a mean density equal to $200$ times the critical density of the universe. 

\subsection{The TNG50 MW/M31-like sample}
We focus our analysis on the sample of MW/M31 analogs in \textsc{TNG50} curated by \citet{Pillepich2023}. 
This sample fulfills the following criteria at $z=0$:
\begin{enumerate}
    \item The stellar mass, $M_{\star}$ (measured within $r \leq 30$ kpc), is in the range $10^{10.5}{\, \rm M_{\odot}}<M_{\rm \star} \leq 10^{11.2}\, {\rm M_{\odot}}$, and the host halo mass is $M_{\rm 200c} < 10^{13}\, {\rm M_{\odot}}$; 
    \item The stellar disc has a minor-to-major axis ratio $c/a \leq 0.45$ (measured between $1-2\, r_{\star, 1/2}$) \textit{or} it appears visually disc-like with obvious spiral arms, and;
    \item No other galaxies with $M_{\star} \geq 10^{10.5}\, {\rm M_{\odot}}$ are within $500$ kpc. 
\end{enumerate}
A total of $198$ disc galaxies satisfy these criteria. More details about their selection can be found in \citet{Pillepich2023}. 

\subsection{The implementation of SMBH in TNG}\label{sec:bhfeedback}
The SMBH model used for TNG is described in \citet{Weinberger2017}. 
In brief, SMBHs are seeded with an initial mass of $M_{\rm BH} = 1.18 \times 10^6\, {\rm M_{\odot}}$ in haloes whose friend-of-friends mass \citep{Davis1985} exceeds $\gtrsim 7.38 \times 10^{10}\, {\rm M_{\odot}}$, provided they do not already contain one. 
SMBHs can grow through mergers or by accreting mass from neighbouring gas cells at the Eddington-limited Bondi accretion rate:
\begin{equation}
    \dot{M}_{\rm BH} = \min \left(\dot{M}_{\rm bondi}, \dot{M}_{\rm edd} \right),
\end{equation}
where
\begin{equation}
    \dot{M}_{\rm bondi} = \frac{4 \pi G^2 M_{\rm BH}^2 \rho_{\rm g}}{c_{\rm s}^3}
\end{equation}
and
\begin{equation}
    \dot{M}_{\rm edd} = \frac{4 \pi G M_{\rm BH} m_{\rm p}}{\sigma_{\rm T} \epsilon_{\rm r} c} 
\end{equation}
are the Bondi-Hoyle–Lyttleton \citep{Hoyle1939, Bondi1944, Bondi1952} and Eddington accretion rates, respectively. 
Here, $G$ is the gravitational constant, $c$ is the vacuum speed of light, $\sigma_{T}$ is the Thompson cross-section, and $m_{\rm p}$ is the proton mass. 
The parameter $\epsilon_{\rm r}=0.2$ is the radiative accretion efficiency, and $\rho_{\rm g}$ and $c_{\rm s}$ are the average density and sound speed of the neighbouring $646$ gas cells. 

Some of the energy accreted is released back into the surrounding gas cells as feedback. 
In \textsc{TNG}, there are two modes of SMBH feedback: (1) at high accretion rates, it is injected as pure thermal energy (hereafter ``thermal mode'' feedback), and (2) at low accretion rates, it is injected as momentum kicks (hereafter ``kinetic mode''). 
The mode of SMBH feedback depends on whether the Bondi-Hoyle–Lyttleton accretion rate exceeds a $M_{\rm BH}$-dependent fraction of the Eddington accretion rate: SMBH feedback is in kinetic mode when 
\begin{equation}\label{eq:feedback_thres}
    \chi = \frac{\dot{M}_{\rm bondi}}{\dot{M}_{\rm edd}} < \chi_{\rm thresh} \equiv \min \left(0.002\left( \frac{M_{\rm BH}}{10^8\, {\rm M_{\odot}}} \right)^{2} , 0.1\right),
\end{equation}
and is in thermal mode otherwise. 
We define the lookback time at which the central SMBH first enters kinetic feedback mode as $t_{\rm kin}$, i.e. the time of the first snapshot at which $\chi < \chi_{\rm thresh}$.

\subsection{Characterising stellar bars in TNG50}\label{sec:baranalysis}
We characterize stellar bars using a Fourier decomposition of the face-on stellar surface mass density \citep[][]{Athanassoula2002, Dehnen2022, Frosst2024}. 
The strength ($A_{\rm m}$) and phase angle ($\phi_{\rm m}$) of a bar is related to the amplitude of the even Fourier modes; they are calculated as follows:
\begin{equation}
    A_{\rm m} = |\mathcal{A}_{\rm m}|
\end{equation}
and
\begin{equation}
    \phi_{\rm m} = \frac{1}{\rm m}\arg (\mathcal{A}_{\rm m}),
\end{equation}
where
\begin{equation}
    \mathcal{A_{\rm m}} = \frac{\sum_{j} M_{j} e^{{\rm m}i\theta_{j}}}{\sum_{j} M_{j}}.
\end{equation}
In the latter, $M_{j}$ and $\theta_{j}$ are the mass and azimuthal angle of the $j^{\rm th}$ stellar particle, and m is the order of the Fourier mode. 
We measure the lowest order bar strength profile, $A_{2}(R)$, in radial bins. 
The bar strength, $A_{2}^{\rm max}$, is defined as the maximum value of $A_{2}(R)$ and has a phase angle defined as $\phi_{2}^{\rm max}$. 
The bar length, $R_{\rm bar}$, is identified as the maximum extent beyond $A_{2}^{\rm max}$ where $\phi_{2}(R)$ deviates from $\phi_{2}^{\rm max}$ by $\leq \pm 10^{\circ}$ while $A_{2}(R) \geq A_{2}^{\rm max}/2$, following the procedure described in \citet{Dehnen2022}. 
We follow \citet{RosasGuevara2022}, and define barred galaxies as those with both $A_{2}^{\rm max} > 0.2$ and $R_{\rm bar} \geq 1.4\epsilon_{\rm c}$, ensuring a constant $\phi_{2}(R)$ in the galaxy centre before bars are identified. 
The bar formation time, $t_{\rm bar}$, is defined as the first time a galaxy is barred and remains so for at least three consecutive snapshots ($\approx 450$ Myr). 

All galaxies are visually inspected at each snapshot to ensure our measurements are not corrupted by transient events like mergers. 
At $z=0$, 125 of the galaxies in our sample have stellar bars, roughly $63$ per cent, in good agreement with observations of galaxies in the local universe and with other studies based on \textsc{TNG50} \citep[e.g.][]{Sheth2008, Nair2010, Masters2011, Gargiulo2022, RosasGuevara2022}. 

\begin{figure}
	\includegraphics[width=\columnwidth]{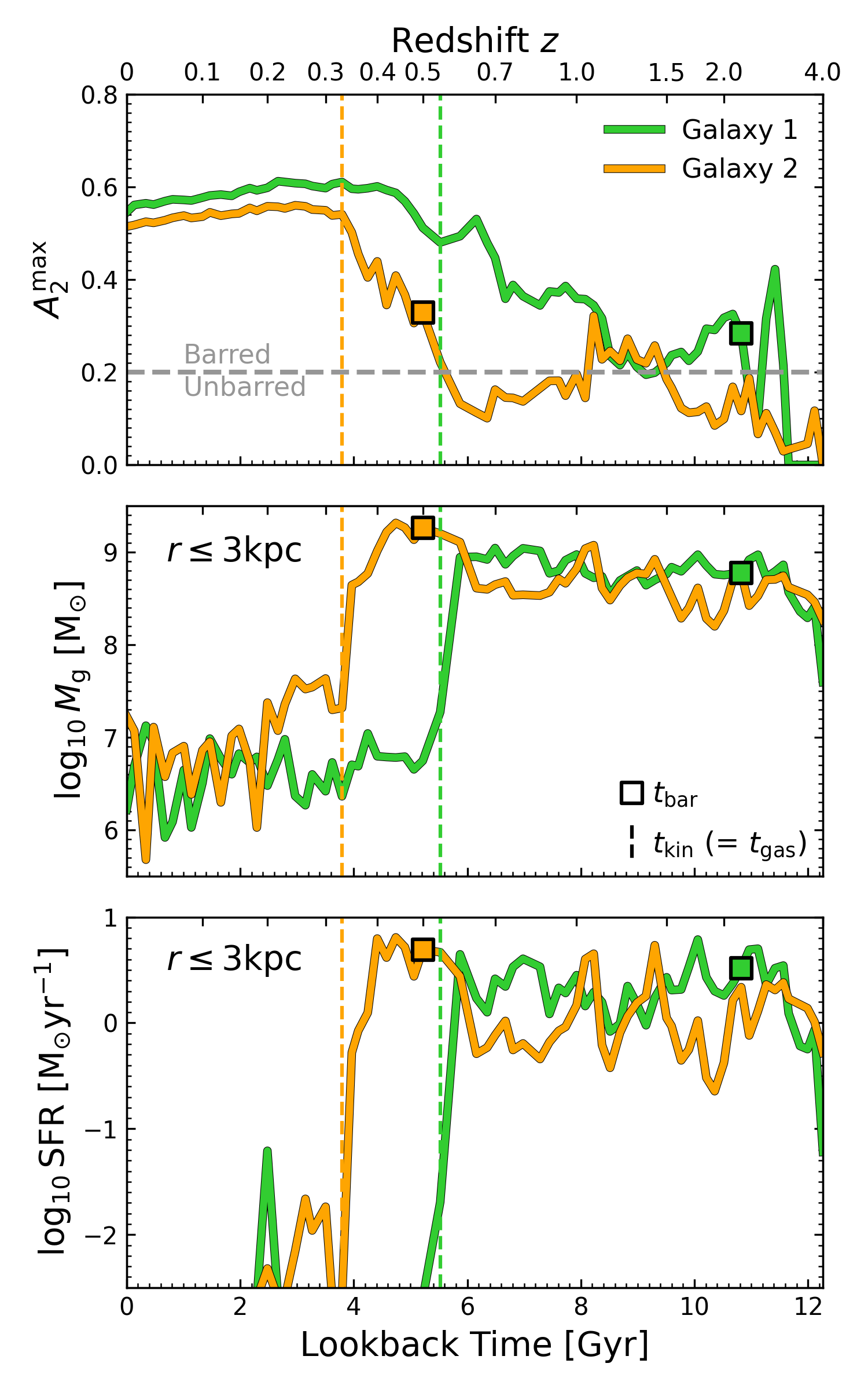}
    \caption{The bar strength, $A_{2}^{\rm max}$ (upper panel), the central gas mass, $M_{\rm g}(r\leq 3\, {\rm kpc}$) (middle panel), and the central SFR within the same radius (lower panel), as a function of time from $z=4$ to $0$. 
    We plot the evolutionary histories of the two example galaxies from Fig.~\ref{fig:projection}, Galaxy 1 and Galaxy 2, as green and orange lines, respectively. 
    The bar formation time, $t_{\rm bar}$, is plotted as a coloured square for each galaxy. 
    The gas hole formation time, $t_{\rm gas}$, and the time at which the central SMBH first enters kinetic feedback mode, $t_{\rm kin}$, occur at the same time and are plotted as coloured vertical dashed lines.
    }
    \label{fig:profile_evolution}
\end{figure}

\section{Links between quenching, SMBHs, and bars}\label{sec:results}
\subsection{A case study of two barred galaxies} \label{sec:morph}
To set the stage, in Fig.~\ref{fig:projection} we plot the stellar and gas face-on surface mass densities for two example galaxies at $z=2$, $1$, $0.5$, and $0$ (left to right, respectively). 
By $z=0$, both galaxies have developed prominent holes in their central gas distributions. 
These holes effectively halt star formation, leaving the inner $3$ kpc largely or entirely quenched: the $z=0$ star formation rates (SFRs) within $r=3\,{\rm kpc}$ are $0\, {\rm M_{\odot}yr^{-1}}$ and $10^{-3.9}\, {\rm M_{\odot}yr^{-1}}$ for Galaxy 1 and 2\footnote{Galaxies 1 and 2 have SubfindIDs 530852 and 534628 at $z=0$, respectively}, respectively \citep[SFRs below $10^{-2.5}\, {\rm M_{\odot}yr^{-1}}$ are considered unresolved and are hence equivalent to no SF, see][]{Donnari2019}. 
By $z=0$, both galaxies host stellar bars, clearly visible in their stellar surface mass density distributions ($A_{2}^{\rm max} = 0.55$ and $0.51$ for Galaxy 1 and 2, respectively), and also have central SMBHs operating in kinetic feedback mode. 

Both the kinetic feedback and bar dynamics could, a priori, contribute to the central quenching of SF. 
For instance, bars may stabilize gas discs against SF \citep[but leave inert gas in the galaxy centre, e.g.][]{Khoperskov2018}, or gas influenced by the bar may trigger nuclear SF and BH accretion, ultimately stifling the central SF once the gas is consumed \citep[e.g.][]{Fanali2015}. 
On the other hand, for reasons unrelated to the bar, SMBH kinetic feedback may eject gas from the galaxy centre, removing the fuel for SF \citep[e.g.][]{Davies2020, Terrazas2020}. 
The clue to distinguishing these mechanisms lies in the evolutionary histories of these galaxies. 

In Fig.~\ref{fig:profile_evolution} we show the time evolution of the stellar bar strength, $A_{2}^{\rm max}$ (upper panel), the total gas mass within $3\, {\rm kpc}$ of each galaxy's centre (middle panel), and the SFR within the same radius (lower panel) for the two example galaxies. The evolution of $A_{2}^{\rm max}$ shows that both galaxies host long-lived bars that persist to $z=0$; the values of $t_{\rm bar}$ are marked along each of the curves using squares. 
Immediately after their bars form, the central gas mass and SFR of each galaxy is largely unchanged: this is particularly evident for Galaxy 1, whose central gas fraction and SFR remains approximately constant for $\approx 5\, {\rm Gyr}$ following $t_{\rm bar}$. 
Eventually, however, both galaxies experience sudden drop ($\approx 1 - 2$ dex) in their central gas mass and SFR, which lasts to $z=0$; this corresponds to the formation of the gas holes already apparent in Fig.~\ref{fig:projection}. 
We mark this drop with a vertical dashed line, and denote it $t_{\rm gas}$. 
Hereafter, $t_{\rm gas}$ is defined as the time at which the gas-to-stellar mass ratio (within $3\, {\rm kpc}$) drops below 1 per cent and remains so for at least three consecutive snapshots ($\approx450$ Myr); as seen in the bottom panel of Fig.~\ref{fig:profile_evolution}, $t_{\rm gas}$ coincides with star formation quenching. 
    
By analyzing the trajectories of tracer particles, we verified that immediately after $t_{\rm gas}$ the vast majority of the central gas mass is ejected into the circumgalactic medium of the galaxies, rather than being consumed by star formation \citep[see][for details]{Genel2013}. 
This suggests that feedback from SMBHs leads to the formation of the central gas holes.
Indeed, $t_{\rm kin}$, i.e. the first snapshot at which the central SMBH is in kinetic mode, exactly coincides with $t_{\rm gas}$ at the time resolution of the simulation and also occurs at the time marked by the vertical dashed lines in Fig.~\ref{fig:profile_evolution}. 
Evidently, both of these galaxies form stellar bars before their gas holes develop, but their gas holes are first identified in the same snapshot that the SMBH enters kinetic feedback mode.

\subsection{Black hole feedback quenches star formation in barred galaxies}\label{sec:quenched}
\begin{figure}
	\includegraphics[width=\columnwidth]{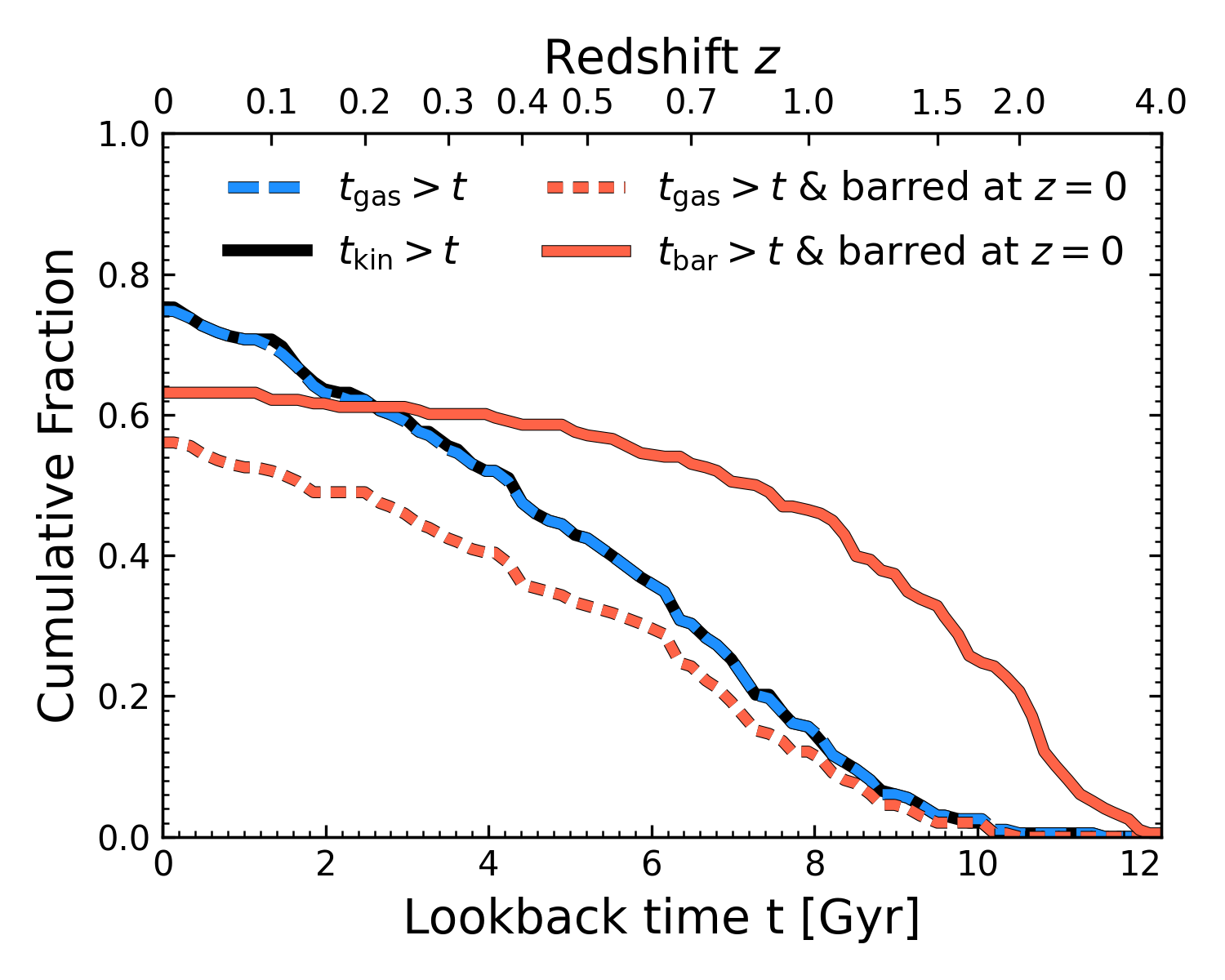}
    \caption{The dashed blue and solid black lines show, respectively, the cumulative fractions of galaxies that have developed gas holes (i.e. the distribution of $t_{\rm gas}$) and those that have central SMBHs in kinetic feedback mode (i.e. the distribution of $t_{\rm kin}$). The solid and dashed red lines show, respectively, the fraction of all galaxies that have developed persistent stellar bars (i.e. the distribution of $t_{\rm bar}$ for discs that host stellar bars at $z=0$) and the subset of barred galaxies that have formed gas holes. Gas holes almost always appear in the same snapshot in which SMBH kinetic feedback is initiated. About 90 per cent of all barred galaxies eventually develop gas holes (but some galaxies with gas holes at $z=0$ are not barred).}
    \label{fig:fraction_evolution}
\end{figure}

In Fig.~\ref{fig:fraction_evolution} we show that the results from the case study in Section~\ref{sec:morph} apply to the whole sample. 
The dashed blue and solid black lines show, respectively, the cumulative fraction of all galaxies that have formed gas holes by a particular time (i.e. the distribution of $t_{\rm gas}$) and those that have entered kinetic feedback mode (i.e. the distribution of $t_{\rm kin}$); they are essentially indistinguishable. 
This indicates that the formation of holes in the central gas distribution of discs coincide with the SMBH entering kinetic feedback mode (we have verified this for each galaxy individually). 
Only about $6$ per cent of galaxies with gas holes have $t_{\rm kin} \neq t_{\rm gas}$. 
We conclude that gas holes form as a result of SMBH kinetic feedback \citep[in agreement with][]{Li2020,Terrazas2020,Zinger2020,Pillepich2021}, and are not caused by stellar bars.

Note also that the majority of galaxies that host stellar bars develop gas holes after their bars form. 
The solid red line in Fig.~\ref{fig:fraction_evolution} shows the distribution of $t_{\rm bar}$ (i.e. the cumulative fraction of all galaxies that have formed a stellar bar by time $t$, and that remain barred until $z=0$). 
Contrast this with the short-dashed red line, which shows the cumulative fraction of the same barred galaxies that have formed gas holes. 
Although these distributions differ, they reveal that about $92$ per cent of all barred galaxies eventually develop central gas holes, and that bars almost always form first. 
Conversely, only around $8$ per cent of barred discs in our sample do not have holes in their central gas distributions at $z=0$. 
This is a non-trivial result that we return to in Section~\ref{sec:halt}.

\begin{figure}
	\includegraphics[width=\columnwidth]{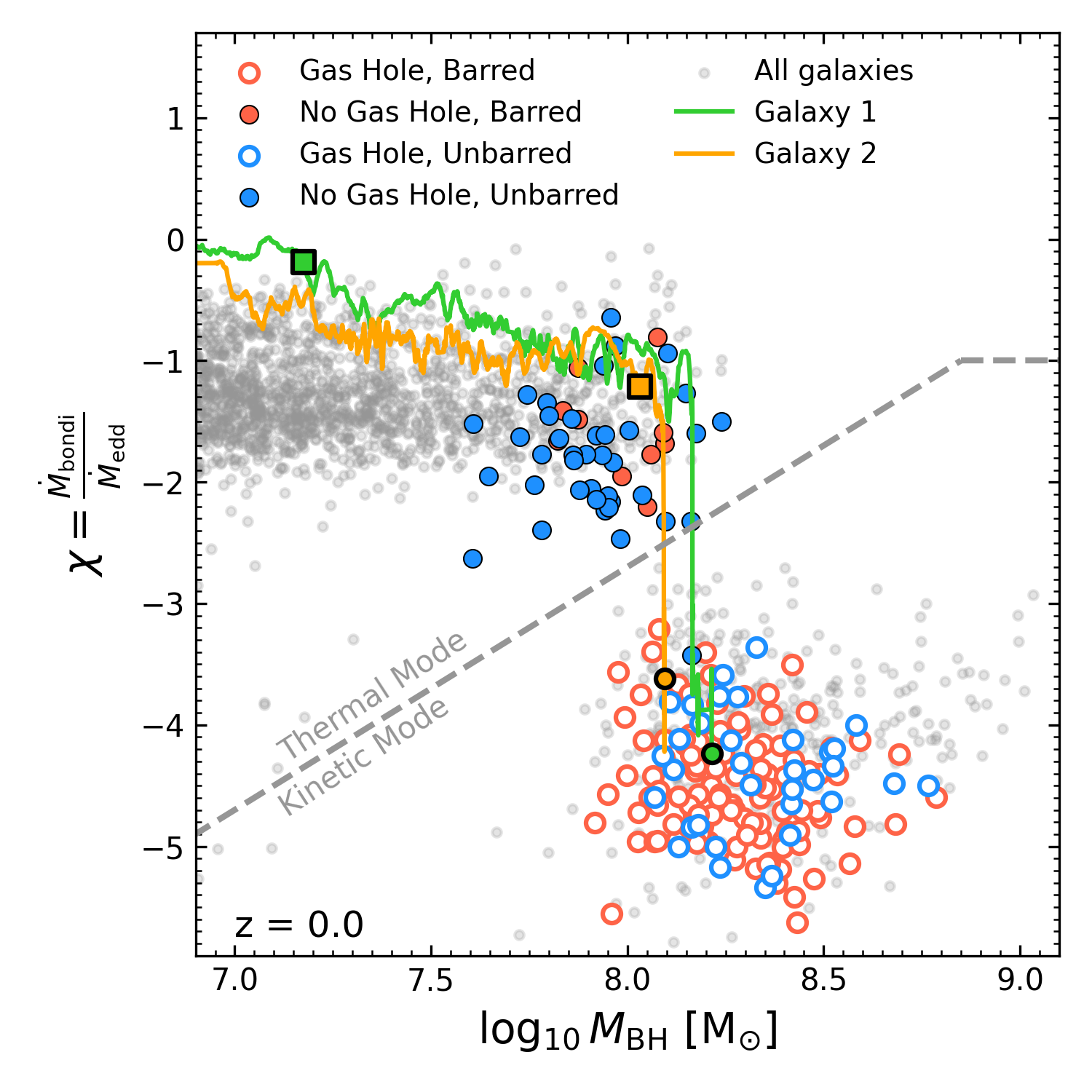}
    \caption{The $z=0$ Bondi-to-Eddington accretion rate ratio, $\chi$, for central SMBHs plotted as a function of black hole mass, $M_{\rm BH}$. The critical accretion ratio that determines whether a black hole is in thermal or kinetic feedback mode is plotted as a grey dashed line (defined by Eq.~\ref{eq:feedback_thres}). Coloured symbols show the sample of discs used in our study, split into those that are barred and unbarred (red and blue symbols, respectively), and that those that do or do not have central gas holes (open and closed symbols, respectively). All other \textsc{TNG50} galaxies are shown as grey points for comparison. Trajectories for the two example galaxies in Figs.~\ref{fig:projection} and \ref{fig:profile_evolution} are plotted as green and orange lines; squares of corresponding colour indicate $t_{\rm bar}$ for these two systems. Galaxies with gas holes have central black holes that operate exclusively in kinetic feedback mode, and many of them also host stellar bars.}
    \label{fig:BHaccretion}
\end{figure}

Given that $t_{\rm gas} \approx t_{\rm kin}$ for almost all of the discs, it is clear that gas holes form due to kinetic winds from the central SMBH, rather than through a bar driven process. 
We confirm this in Fig.~\ref{fig:BHaccretion}, where we plot the $z=0$ Bondi-to-Eddington accretion rate ratio, $\chi$, versus $M_{\rm BH}$. 
Discs are split into barred and unbarred (red and blue points, respectively), and galaxies with and without gas holes (open and closed circles, respectively). 
For comparison, the grey points show all other SMBHs in \textsc{TNG50} (no distinction has been made between centrals and satellites), and coloured lines show the trajectories of Galaxy 1 and 2 from Fig.~\ref{fig:projection}. 
The separation between galaxies whose SMBH are in thermal versus kinetic feedback mode is shown as a dashed grey line (defined by eq.~\ref{eq:feedback_thres} in Section~\ref{sec:bhfeedback}). 

Fig.~\ref{fig:BHaccretion} elicits some comments. First, two distinct black hole populations are evident: those in thermal feedback mode (high $\chi$, $M_{\rm BH}\lesssim 10^8\,{\rm M_\odot}$), and those in kinetic feedback mode (low $\chi$, $M_{\rm BH}\gtrsim 10^8\,{\rm M_\odot}$). 
Second, all galaxies with holes in the centres of their gas discs at $z=0$ have SMBHs operating in kinetic feedback mode (i.e., the open circles are below the grey dashed line), and most, but not all, of these galaxies are barred (about $77$ per cent of discs in kinetic mode at $z=0$ are barred; red open circles). This clearly shows that the gas holes (and subsequent quenching) are driven by SMBH kinetic winds rather than bar driven quenching. Similar gas holes have been found in other galaxy formation simulations that employ two-mode SMBH feedback \citep[e.g.][]{Irodotou2022, Wellons2023, Arora2024}. However, this does not explain why most discs in our sample that contain central gas holes are also barred. 

The green and orange lines in Fig.~\ref{fig:BHaccretion} show that the example galaxies above initially host low-mass SMBHs operating in thermal feedback mode. As their central SMBHs grow, they later enter kinetic mode feedback when $M_{\rm BH} \gtrsim 10^{8}\, {\rm M_{\odot}}$. Furthermore, the early bar formation in Galaxy 1 (green line) does not significantly affect the evolutionary history of its SMBH during the nearly $5$ Gyr between $t_{\rm bar}$ and $t_{\rm kin}$. 
When $t_{\rm bar} \gg t_{\rm kin}$ (green line), the SMBH has relatively low mass at $t_{\rm bar}$ and is operating in thermal mode for much of the galaxy's history, but if $t_{\rm bar} \approx t_{\rm kin}$ (orange line), the galaxy forms a bar when the SMBH is already massive and about to enter kinetic feedback mode. 

\begin{figure}
	\includegraphics[width=\columnwidth]{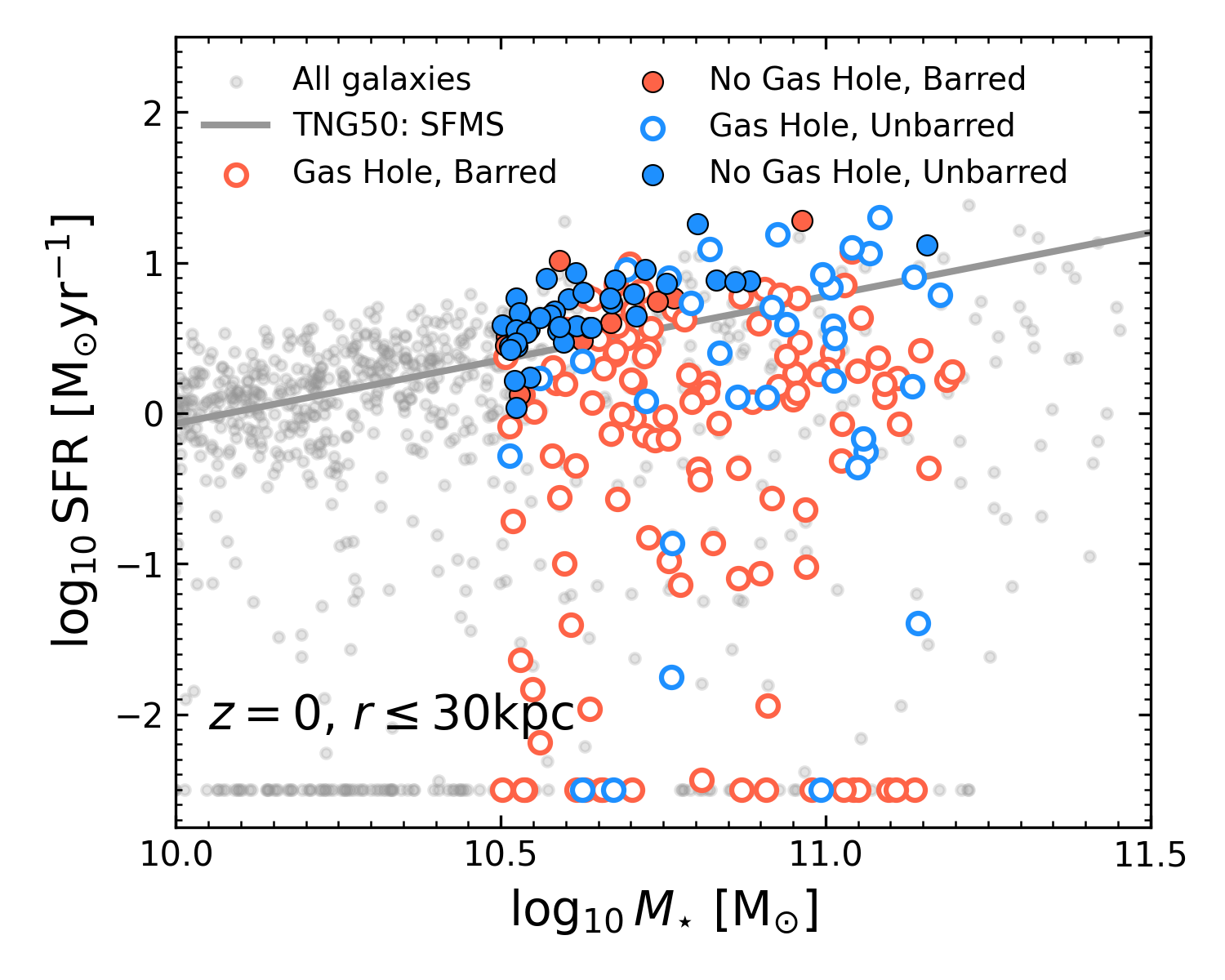}
    \caption{The $z=0$ SFR plotted versus galaxy stellar mass; both quantities were measured within an aperture of $r = 30$ kpc. Coloured symbols follow the plotting conventions used for Fig.~\ref{fig:BHaccretion}. The presence of a gas hole driven by SMBH kinetic feedback clearly affects a galaxy's star formation rate. Whether a galaxy hosts a bar or not has no effect. For comparison, we plot the star formation main sequence (SFMS) for \textsc{TNG50} as a solid grey line; grey points show individual galaxies not in our sample. The band of galaxies with ${\rm SFR}=10^{-2.5}\,{\rm M_\odot yr^{-1}}$ have unresolved SFRs.}
    \label{fig:sfms_radii}
\end{figure}

In Fig.~\ref{fig:sfms_radii} we plot the $z=0$ instantaneous SFR of gas cells versus $M_{\star}$ (both quantities were measured within $r = 30$ kpc). For comparison, we plot the star formation main sequence (SFMS) for \textsc{TNG50} as a solid grey line \citep{Donnari2019} and the remaining \textsc{TNG50} galaxies as grey points. 
As in Fig.~\ref{fig:BHaccretion}, we distinguish barred and unbarred galaxies (red and blue points, respectively) and galaxies with and without gas holes (open and closed circles, respectively). 
Galaxies with gas holes (which also have SMBH in kinetic feedback mode) have lower SFRs than galaxies without gas holes. 
This is true regardless of whether the galaxy is barred or not. 
For our sample of \textsc{TNG50} discs, SMBH feedback is an efficient quenching mechanism, whereas stellar bars do not influence the SFR either globally or locally. 

\subsection{Secular bar formation is halted in galactic discs that have been quenched by black hole feedback}\label{sec:halt}
We have established that the gas holes present in many of the discs in our sample are created by kinetic BH feedback and not by stellar bars. Nevertheless, galaxies with prominent gas holes are preferentially barred, and vice versa. We next explore the relation between the presence of stellar bars and gas holes. 

\begin{figure}
	\includegraphics[width=\columnwidth]{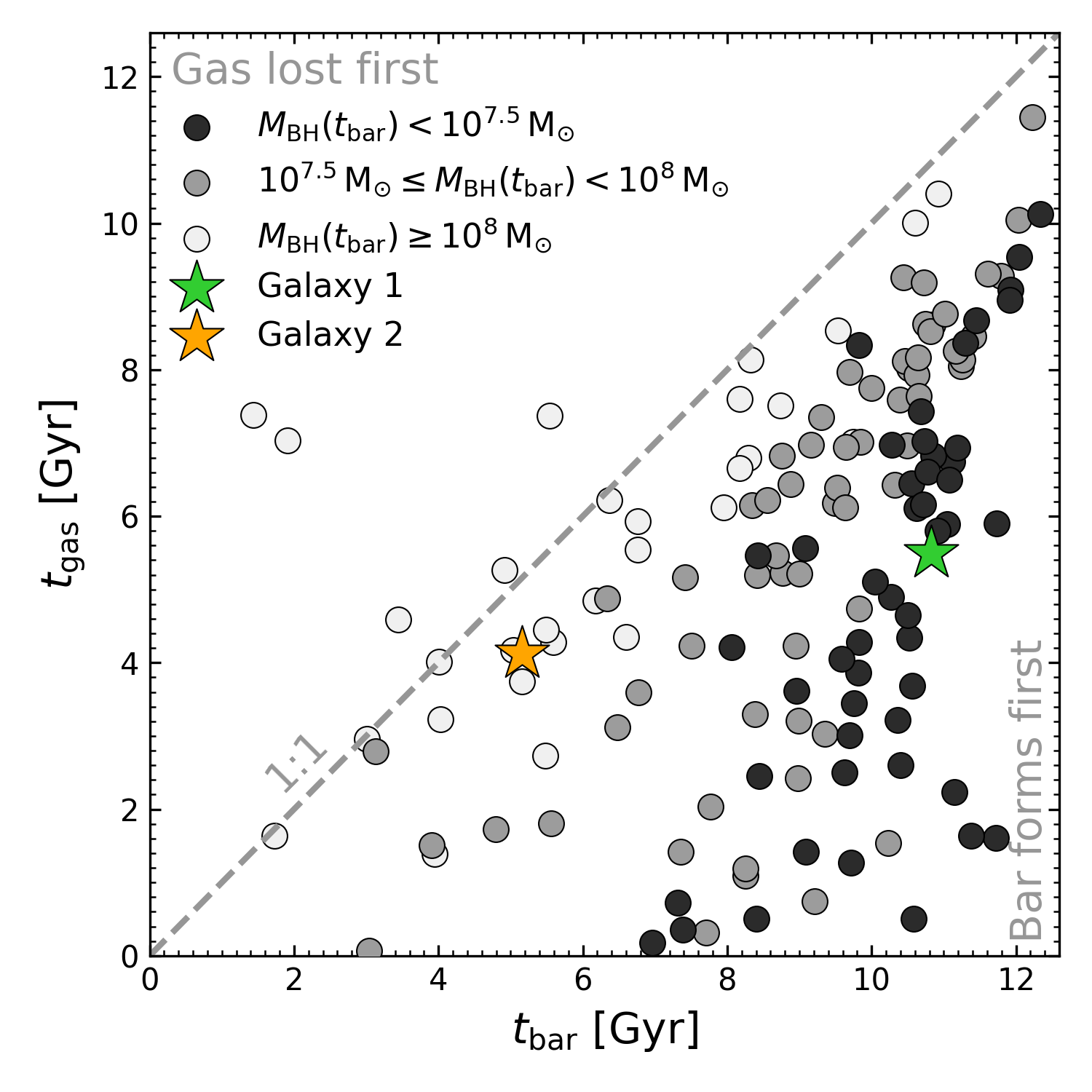}
    \caption{The lookback time at which gas holes form, $t_{\rm gas}$, plotted versus the bar formation time, $t_{\rm bar}$. Points are shaded according to $M_{\rm BH}(t_{\rm bar})$, i.e. the mass of the central SMBH at $t_{\rm bar}$. The two example galaxies from Fig.~\ref{fig:projection} are shown as green and orange stars, as labeled. Note that for the vast majority of galaxies $t_{\rm bar} \gtrsim t_{\rm gas}$, and that the relation between them differs depending on the mass of the central black hole at $t_{\rm bar}$: galaxies with $M_{\rm BH}(t_{\rm bar})\approx 10^8\,{\rm M_\odot}$ (the critical mass for kinetic BH feeback; see Fig.~\ref{fig:BHaccretion}) have $t_{\rm gas}\approx t_{\rm bar}$, whereas those with $M_{\rm BH}(t_{\rm bar}) < 10^{7.5}\,{\rm M_\odot}$ have $t_{\rm gas} \ll t_{\rm bar}$. Galaxies in our sample rarely develop stellar bars after $t_{\rm gas}$. }
    \label{fig:distribution}
\end{figure}

In Fig.~\ref{fig:distribution} we plot the lookback times at which galaxies develop their gas holes ($t_{\rm gas}$) versus the time at which they form their stellar bars ($t_{\rm bar}$); note that only barred galaxies that also contain gas holes are plotted. 
This comparison reveals that bars almost always form \textit{before} gas holes do, as expected from Fig.~\ref{fig:fraction_evolution}. 
Only eight galaxies form a stellar bar after a gas hole has already formed; visual inspection suggests that most of these eight bars were triggered by a passing satellite galaxy. 

To highlight how these timescales relate to SMBH feedback, we shaded the points in Fig.~\ref{fig:distribution} by the mass of the galaxy's central SMBH at $t_{\rm bar}$, i.e. by ${\rm M}_{\rm BH}(t_{\rm bar})$.  Galaxies with the largest time delay between $t_{\rm bar}$ and $t_{\rm gas}$ also have the lowest-mass SMBHs at $t_{\rm bar}$ (represented by black points) -- their bars form long before their SMBHs are massive enough to enter kinetic feedback mode. On the other hand, galaxies with ${\rm M_{BH}}(t_{\rm bar})\geq 10^8\,{\rm M_\odot}$ tend to develop central gas holes shortly after their bars form (i.e. the white points have $t_{\rm gas}\approx t_{\rm bar}$). This is because, in the TNG model, SMBHs enter kinetic mode when they reach a critical mass ${\rm M}_{\rm BH} \gtrsim 10^{8}\, {\rm M_{\odot}}$ \citep[see Fig.~\ref{fig:BHaccretion}, and][]{Terrazas2020, Zinger2020}. But why are galaxies apparently unable to form stellar bars {\em after} their SMBHs enter kinetic mode?

\begin{figure}
	\includegraphics[width=\columnwidth]{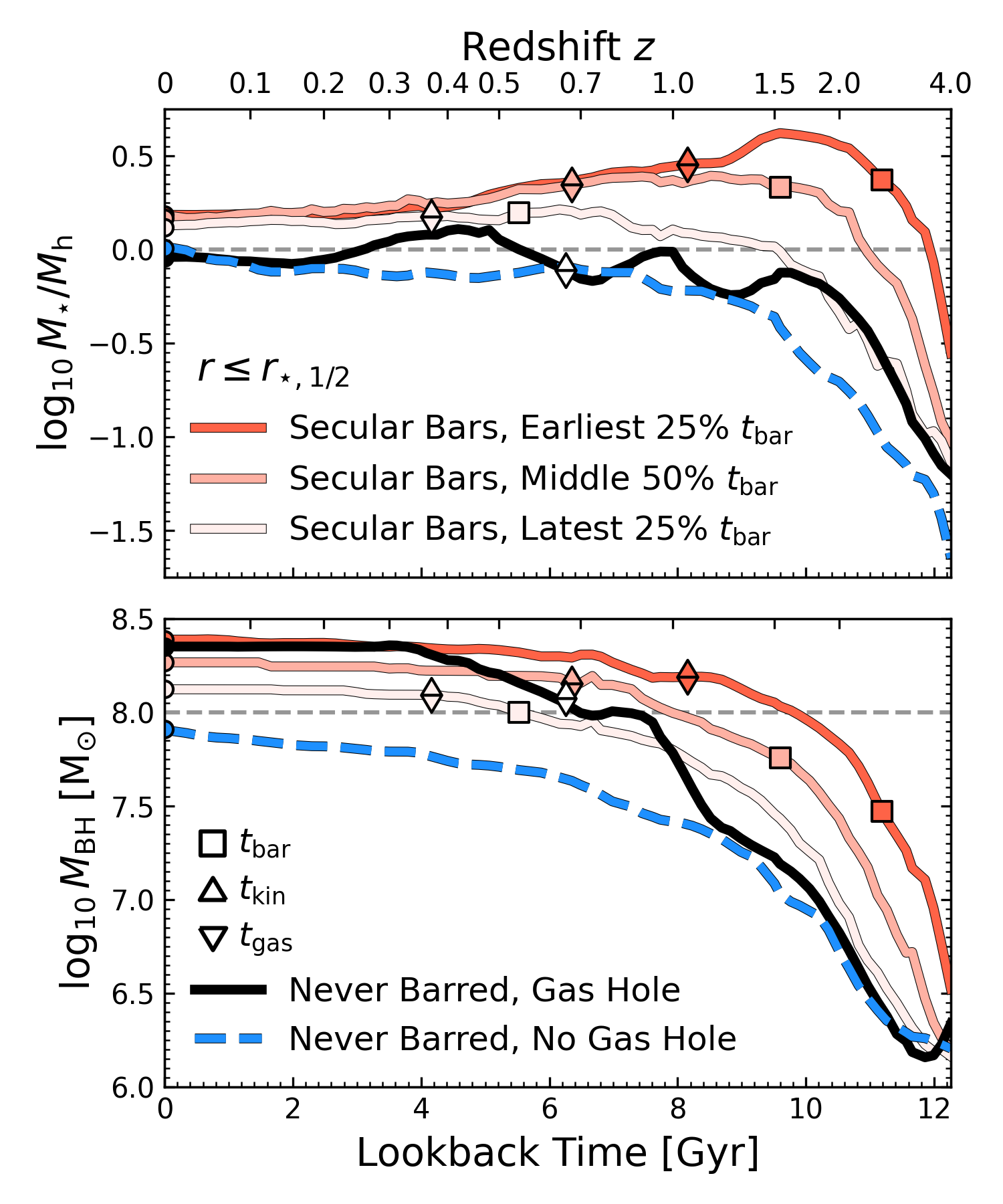}
    \caption{The median stellar-to-dark matter mass ratio (measured within $r_{\star,1/2}$; upper panel) and median SMBH mass (lower panel) as a function of lookback time. Lines with a red hue correspond to galaxies that form stellar bars secularly; they are divided into separate bins based on the relative time at which their bars formed (as indicated in the legend). The blue dashed line corresponds to the sample of galaxies that never formed bars, and the black line shows galaxies that never formed bars but did form gas holes. The horizontal grey dashed line in the upper panel separates galaxies that are disc-dominated (${\rm M_{\star}/M_{\rm h}} \geq 1$) and dark matter-dominated (${\rm M_{\star}/M_{\rm h}} < 1$). In the lower panel, the horizontal grey dashed line correspond to ${\rm M_{\rm BH} = 10^8\, {\rm M_{\odot}}}$, roughly the critical mass for kinetic SMBH feedback. Note that disc galaxies will only develop stellar bars after they become disc-dominated, and they only develop gas holes provided ${\rm M_{\rm BH} \gtrsim 10^8\, {\rm M_{\odot}}}$}
    \label{fig:massfrac}
\end{figure}

In the upper panels of Fig.~\ref{fig:massfrac} we plot the time evolution of the median stellar-to-dark matter mass ratio (measured within the evolving stellar half-mass radius, $r_{\star, 1/2}$) for several subsets of our disc sample. 
In the lower panel we plot the evolution of the median ${\rm M_{\rm BH}}$ for the same galaxy samples. The various samples comprise galaxies that have never formed bars, and those that form bars secularly.\footnote{Galaxies that host secular bars are those that have not experienced a merger with a stellar mass ratio $\mu \geq 0.1$ within $500$ Myr of $t_{\rm bar}$ \citep[see][for details about how these mergers are defined]{Bottrell2024}. Importantly, we determine $\mu$ by evaluating it at the snapshot at which the mass of the least massive progenitor the is greatest \citep[e.g.][]{SotilloRamos2022}. Using merger time windows that are reasonably larger or smaller than 500 Myr does not affect our results.} 
Discs that were never barred are plotted using dashed blue lines if they do not possess a gas hole and as solid black lines if they do possess a gas hole. 
The galaxies that form bars secularly -- almost all of which also possess gas holes -- were split into three bins of $t_{\rm bar}$: those in the upper quartile of $t_{\rm bar}$ (red lines), those in the lower quartile of $t_{\rm bar}$ (light pink lines), and those in between. 
Along each of these curves, we plot the median values of $t_{\rm bar}$, $t_{\rm gas}$, and $t_{\rm kin}$ as squares, upward triangles, and downward triangles, respectively. 
 
The curves plotted in the upper panel of Fig.~\ref{fig:massfrac} show the median stellar-to-dark matter mass ratio (within $r_{\star, 1/2}$) of barred discs exceeds that of unbarred discs at all times. 
Note too that barred galaxies become disc dominated (${\rm M_{\star}/M_{\rm h}}>1$) before their bars form, and those that form their bars earlier also become disc dominated earlier \citep[see also][]{Kraljic2012, Sheth2012}. 
In contrast, galaxies that never form bars are dark matter-dominated within their stellar half-mass radii at all times. 
Having a gravitationally-dominant disc appears to be a prerequisite for bar formation \citep[see also][]{RosasGuevara2020, Lopez2024, Fragkoudi2024}. 
This agrees with the early analytic work of \citet{OP1973}, who showed that strongly self-gravitating discs are susceptible to bar formation.

The lower panel of Fig.~\ref{fig:massfrac} shows that galaxies that become disc dominated and form secular bars earlier also tend to have central SMBH that grow more rapidly at early times \citep[see also][]{RosasGuevara2020, Zhou2020}. 
These early growing SMBHs eventually reach sufficiently high masses ($M_{\rm BH} \gtrsim 10^8\, {\rm M_{\odot}}$) to drive kinetic winds that quench their galaxies, but this usually occurs after the galaxies have already become disc-dominant and therefore susceptible to bar formation. 
This is the case for the barred galaxy samples represented by red lines in Fig.~\ref{fig:massfrac}; the downward triangles along these lines mark $t_{\rm kin}$, which occurs long after ${\rm M_\star/M_h}> 1$. 

However, for a small number of galaxies, the SMBHs will grow sufficiently massive to drive kinetic winds and quench the galaxy {\em before} the disc's gravity dominates that of the halo. 
These galaxies (represented by the black lines in Fig.~\ref{fig:massfrac}) do not form bars because the disc was unable to become sufficiently massive to enable bar formation before being quenched: at early times, these galaxies have similar stellar-to-dark matter mass fractions to those hosting late-forming secular bars (light pink line), but after kinetic feedback is triggered by a period of exceptionally fast black hole growth, their stellar-to-dark matter mass ratios level off and they remain dark matter-dominated thereafter. 
Indeed, for all galaxies in our sample, $M_{\star}/M_{\rm h}$ remains constant or declines after they are subject to strong SMBHs feedback. 
On the other hand, the late forming galaxies represented by the blue dashed line do not have bars -- because they were never disc dominated -- or gas holes -- because ${\rm M_{BH}}$ never crossed the critical threshold for kinetic feedback. 
The high fraction of barred galaxies with gas holes and the fact that gas holes tend to form after stellar bars both have their origins in the co-evolution of the stellar disc mass and the mass of the central SMBH.

\section{Summary and Conclusions}\label{sec:conclusions}
We used 198 isolated disc galaxies from \textsc{TNG50} that span the mass range $10^{10.5} \leq {\rm M_\star/M_\odot}\leq 10^{11.2}$ \citep[see][for details]{Pillepich2023} to study the connection between supermassive black hole feedback, star formation quenching, the morphology of gas discs, and the formation of stellar bars. Our main conclusions are as follows:

\begin{enumerate}
    \item By $z=0$, many ($63$ per cent) of the galaxies in our sample have developed bars in their stellar discs and/or $3 - 15$ kpc wide "holes" in their gas discs ($75$ per cent). The formation of central gas holes quenches the central star formation rates (Fig.~\ref{fig:profile_evolution}). Intriguingly, almost all barred galaxies possess gas holes, and in almost every case the bar forms prior to the formation of the gas hole (Fig.~\ref{fig:fraction_evolution}). 
    At first glance, this seems to imply that bars may be responsible for the formation of gas holes and the subsequent quenching of star formation, as is occasionally suggested by observations \citep[e.g.][]{Masters2011, George2019, FraserMcKelvie2020a, Newnham2020}. 
    
    \item However, in Section~\ref{sec:quenched} we showed that almost all galaxies whose centers are gas poor have also been subject to strong kinetic feedback from their central supermassive black holes (SMBHs), and that the gas holes appear immediately after the SMBHs enter kinetic feedback mode (see Fig.~\ref{fig:fraction_evolution}), occurring when the SMBH mass exceeds ${\rm M_{BH}\gtrsim 10^8\, M_\odot}$ (Fig.~\ref{fig:BHaccretion}). 
    This suggests that, for \textsc{TNG50}, kinetic winds from SMBHs are sufficiently strong to eject gas from the centres of discs, overpowering any stellar bar-driven quenching processes (Fig.~\ref{fig:sfms_radii}). Support for this conclusion is provided by a subset of galaxies (roughly $17$ per cent of the total sample) whose stellar discs are not barred but whose gaseous discs have nonetheless developed central holes as soon as the SMBHs turn to kinetic feedback. 

    \item Galaxies whose SMBHs enter kinetic feedback mode are largely quenched of star formation thereafter (Fig.~\ref{fig:sfms_radii}). This explains why the formation of stellar bars always precedes the formation of gas holes (Fig.~\ref{fig:distribution}): if galaxies are not massive enough to support the growth of bar instabilities before being quenched by SMBH feedback, their stellar discs will remain gravitationally subdominant thereafter, and thus stable to bar formation (Fig.~\ref{fig:massfrac}). Exceptions to this (i.e. when bars form after gas holes) appear possible, for example when bars are triggered at externally through tidal encounters with passing satellite galaxies. We plan to address the latter point in future work using a larger sample of disc galaxies that are not subject to the strong isolation criterion imposed on the sample studied here.
    
    \item The high fraction of barred galaxies with gas holes has its origin in the co-evolution of the stellar disc mass and the mass of the central SMBH. Galaxies that form massive discs at early times also form massive BHs at early times, and are thus susceptible to both secular bar formation -- which occurs when the stellar mass exceeds the dark matter mass within $r_{\star,1/2}$ -- and to kinetic SMBH feedback (Fig.~\ref{fig:massfrac}).  
    
\end{enumerate}

An important implication of our study is that fine-tuned models for SMBH feedback can have significant effects on the properties of stellar bars (such as their ages, and the galaxies in which they are able to form) and on the inner morphology and star formation rates of gas discs. A note of caution, however, is that our analysis focused on a set of isolated disc galaxies that span a relatively narrow range of halo masses ($10^{11.5} \lesssim {\rm M_{200c}/M_\odot} \lesssim 10^{12.8}$), which is roughly where SMBH feedback and bar formation are common. Our analysis was also based on only one simulation, {\textsc{TNG50}}, for which kinetic BH feedback is particularly efficient at quenching galaxies \citep[e.g.][]{Terrazas2020}. Future work should extend our analysis to a broader range of galaxy masses and environments, using hydrodynamical simulations that employ different models for SMBH feedback. While real \HI gas holes are observed in some galaxies \citep[e.g.][]{George2019, Newnham2020, Murugeshan2019}, this does not mean that all of the gas is removed or subject to phase change. Careful comparison of the incidence of stellar bars and the morphology of gas discs found in simulations and in observations can possibly be used to constrain subgrid models for SMBH feedback.

\section*{Acknowledgements}
We thank Lars Hernquist and Amelia Fraser-McKelvie for useful conversations. ADL and DO acknowledge financial support from the Australian Research Council through their Future Fellowship scheme (project numbers FT160100250, FT190100083, respectively). CB gratefully acknowledges support from the Forrest Research Foundation. This research was undertaken with the assistance of resources and services from the National Computational Infrastructure (NCI), which is supported by the Australian Government. The IllustrisTNG simulations were undertaken with compute time awarded by the Gauss Centre for Supercomputing (GCS) under GCS Large-Scale Projects GCS-ILLU and GCS-DWAR on the GCS share of the supercomputer Hazel Hen at the High Performance Computing Center Stuttgart (HLRS), as well as on the machines of the Max Planck Computing and Data Facility (MPCDF) in Garching, Germany. This work has benefited from the following public \textsc{Python} packages: \textsc{Scipy} \citep{Virtanen2020}, \textsc{Numpy} \citep{Harris2020}, and \textsc{Matplotlib} \citep{Hunter2007}.

\section*{Data Availability}

The TNG50 simulation data (and MW/M31-like sample) are publicly available; see \citet{Nelson2019a}, \citet{Pillepich2018a}, and \citet{Pillepich2023} for further information. 
Additional data can be made available upon reasonable request. 
 


\bibliographystyle{mnras}
\bibliography{references} 


\appendix

\bsp	
\label{lastpage}
\end{document}